\documentclass[twocolumn,twoside,slac_two]{revtex4}
\usepackage{graphicx}
\usepackage{fancyhdr}
\usepackage{float}
\begin{document}

\title{Time and Space Dependent Stochastic Acceleration Model for the Fermi Bubbles}

%

\author{K. Sasaki, K. Asano, T. Terasawa}
\affiliation{University of Tokyo/Institute for the Cosmic Ray Research, Kashiwa, Chiba, Japan}
\begin{abstract}
Fermi-LAT reveals two huge gamma-ray bubbles existing in the Galactic Center, called 'Fermi Bubbles'. The existence of two microwave bubbles at the same region are also reported by the observation by WMAP, dubbed 'WMAP haze'. In order to explain these components, It has been argued that the gamma-rays arise from Inverse-Compton scattering of relativistic electrons accelerated by plasma turbulence, and the microwaves are radiated by synchrotron radiation. But no previous research reproduces both the Fermi Bubbles and WMAP haze under typical magnetic fields in the galaxy.
We assume that shocks present in the bubbles and the efficiency of the acceleration by plasma turbulence, 'stochastic acceleration', changes with the distance from the shock front. The distance from the shock front increases with time, accordingly the efficiency of the acceleration changes with time. We also consider the time development of the electrons escape from the turbulence by diffusive loss. Our model succeed to reproduce both the observed characteristics of the Fermi Bubbles and WMAP haze under typical magnetic fields.      
\end{abstract}

\maketitle

\thispagestyle{fancy}

\section{Introduction}
\label{sec:intro}
Gamma ray data from the Fermi-LAT reveal two giant bubbles extending up to $\sim50^{\circ}$ above and below the Galactic disk, called 'the Fermi Bubbles(FBs)'\cite{su10}\cite{ack14}. The spectra of the FBs  are harder than ambient radiation fields, and they have 'sharp edge', which means sudden change of the brightness at the boundary. The surface brightness is almost constant in the FBs, though it doesn't mean the FBs have constant volume emissivities. There also exists the microwave bubbles in the same region, dubbed 'the WMAP haze'\cite{dob08}. These huge structures may suggest past large-scaled activities in the Galactic Center(GC), near the massive black hole Sgr $A^{\ast}$. 

Two representative models have been argued in order to explain both the FBs and the WMAP haze. 'Hadronic model' reproduces the FBs by the gamma ray from the $\pi^0$-decay process, and the WMAP haze by the radiation from the secondary electrons\cite{cro11}\cite{fuj13}. In the Hadronic model, protons are accelerated by the Diffusive Shock Acceleration(DSA) or other process, and they make $\pi^0$ after collision with cosmic-ray protons. $\pi^{\pm}$ are also made in this process, and they decay to $e^{\pm}$. These secondary particles radiate microwave in the magnetic field by the synchrotron radiation. 

In contrast, 'Leptonic model' explains the FBs by the gamma ray from the Inverse Compton scattering(IC) process, and the WMAP haze by the synchrotron radiation from the relativistic electrons\cite{che11}\cite{mer11}. In the Leptonic model, it is assumed that electrons are accelerated by the second order Fermi acceleration(stochastic acceleration) by plasma turbulence or molecular clouds, in order to explain hard spectra of the FBs. These relativistic electrons give their energy to the ambient photons by the IC process, and they are observed as gamma ray. Some electrons radiate microwaves by the synchrotron radiation, and the WMAP haze are made.

However, both Hadronic and Leptonic model have problems. Hadronic models require so much energy $\sim10^{57}$ erg, and estimated microwave spectrum conflicts with the WMAP data\cite{fuj13}. In the leptonic models, it is difficult to explain the WMAP haze under typical galactic magnetic field($B\sim4\:\mu$G), so it is required strong magnetic field in the large scale\cite{mer11}.

So we construct the extended model based on the stochastic acceleration model by Mertsch \& Sarkar\cite{mer11}. Their model assumes that electrons are accelerated by the turbulence whose acceleration efficiency decreases with the distance from the shock front. But they don't consider the time development of the electron spectrum, and presume that electrons are at steady states. They also neglect the contribution of the electrons escape from the turbulence. So we extend their model by considering both time development and electrons out of the turbulence. In the sec.\ref{sec:models}, we explain our acceleration model.
And in the sec.\ref{sec:result}, we show the results of our numerical calculation.
\section{Models}
\label{sec:models}

We assume thet there is the shock front near the edge of the FBs and the plasma turbulences are drifted from the shock front with time by the advection at $V=v_{\rm pro}$.

In order to calculate the electron spectrum, we solve the Fokker-Planck equation, 
\begin{equation}
	\frac{\partial n}{\partial{t}}- \frac{\partial}{\partial p}\left(p^2D_{pp}\frac{\partial}{\partial p}\frac{n}{p^2} \right) + \frac{n}{t_{esc}}+\frac{\partial}{\partial p}\left( \frac{dp}{dt}n\right)- Q_{inj}= 0
\label{FPeq}
\end{equation}

where the momentum diffusion coefficient $D_{pp}$ in the second term is described\cite{mer11}

\begin{equation}
	D_{pp}(\xi) = p^2 \frac{8\pi D_{xx}(\xi)}{9}\int_{1/L}^{k_d(\xi)}dk \frac{k^4W(k,\xi)}{v_{\rm F}^2(\xi)+D_{xx}^2(\xi)k^2}
\label{Dpp_xi}
\end{equation}
which represents the farther from the shock, the weaker the acceleration efficiency becomes.

In (\ref{Dpp_xi}) $p$ is energy, $k$ is wavenumber, $v_{\rm F}(\xi)$ is the velocity of the fast-mode wave, and $W(k,\xi)=(u^2/4\pi)L^{-2/3}k^{-11/3}$ is the energy density of the Kolmogorov turbulence. $D_{xx}(\xi)$ is the spatial diffusion coefficient.
The dimensionless parameter $\xi \equiv x/L$($x$ is the distance from the shock front, and $L$ is the turbulence size) represents how far from the shock front.

The third term describes the diffusive loss from the turbulence, and then $t_{esc}$ is accounted for $t_{esc}=L^2/D_{xx}$.

The forth term of (\ref{Dpp_xi}) corresponds to the energy loss of the synchrotron radiation and the IC scatering,

\begin{equation}
	\left( \frac{dp}{dt}\right)_{\rm cool} = P_{\rm syn} + P_{\rm IC}
\end{equation}
\begin{eqnarray}
	P_{\rm syn} &=& \frac{4}{3} c \sigma_{\rm T} \gamma^2 \beta^2 U_B \\
	P_{\rm IC} &=& \frac{4}{3} c \sigma_{\rm T} \gamma^2 \beta^2 U_{ph}\times f_{\rm KN}
\end{eqnarray}

\begin{figure}[H]
\centering
\includegraphics[bb = 50 50 550 290,width=12.0cm]{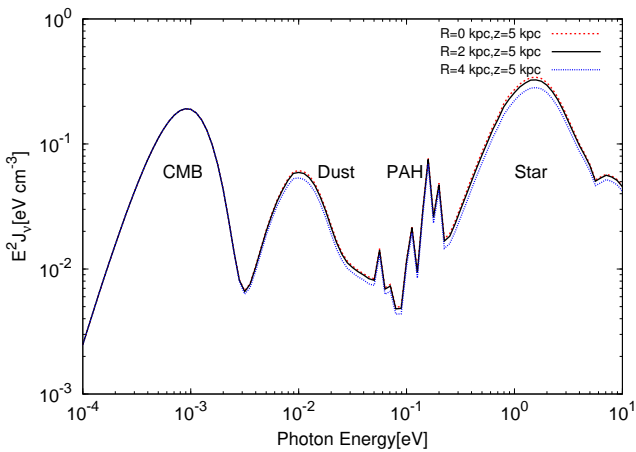}
\caption{Background Photon Fields\cite{vla11}.We use the black solid line($R=2$ kpc,$z=5$ kpc) as $U_{ph}$.}
\label{bgph}
\end{figure}

 $U_B = B^2/8\pi$ where $B = 4 \: \mu G$, and $U_{ph}$ is shown in Figure~\ref{bgph}. $f_{\rm KN} \leq 1$ is the factor due to the Klein--Nishina effect,which is numerically evaluated.
In our models, $v_{\rm pro}$ is fixed $v_{\rm pro} = 250 {\rm \: km \: sec^{-1}}$, which corresponds to sound speed at $kT \sim$ keV. 

We presume that injection rate is constant,
\begin{equation}
	Q_{inj}(p) \equiv Q_0\:\delta(p-p_0)
\end{equation}
where the spectrum shape of injected electron is assumed to be $\delta$-function, $n_{inj}(p)\propto \delta(p-p_0)$. In this paper,we use $cp_0=10^8$ eV.

We consider the effect of time development, so we assume that the distance from the shock front changes with time like $\xi=x/L=v_{\rm pro}t/L$, and then $D_{pp}$ becomes the function of the time $t$. We also consider the contribution of the electrons out of the turbulence, so we solve the equation of these electrons
\begin{equation}
	\frac{\partial n_{esc}}{\partial{t}}+\frac{\partial}{\partial p}\left( \frac{dp}{dt}n_{esc}\right)- Q^{esc}_{inj}= 0
\label{FPesc}
\end{equation}
where $Q^{esc}_{inj}$ is equal to the escape rate from the turbulence $Q^{esc}_{inj}=n/t_{esc}$.

To verify the effect of the escape, we calculate the spectrum of electrons under three cases. 'No escape model' is the model in which we neglect the escape term $n/t_{esc}$(the limit $t_{esc}\rightarrow \infty$). In the 'Cut escape model', we consider the effect of escape but we don't deal with electron out of the turbulence(like\cite{mer11}). So in the Cut escape model, we solve only (\ref{FPeq}), and don't solve (\ref{FPesc}). We consider the electrons out of the turbulence in the 'With escape model', then we solve both (\ref{FPeq}) and (\ref{FPesc}), and finally calculate sum of them $n_e = n+n_{esc}$.

After the calculation of the spectrum of the electrons, we estimate the Intensity from those electrons and compare with the data of the Fermi-LAT and the WMAP. In the next section  sec.\ref{sec:result}, we show the results of our calculation. 
\section{Results}
\label{sec:result}

\begin{figure*}[t]
\centering
\includegraphics[width=180mm]{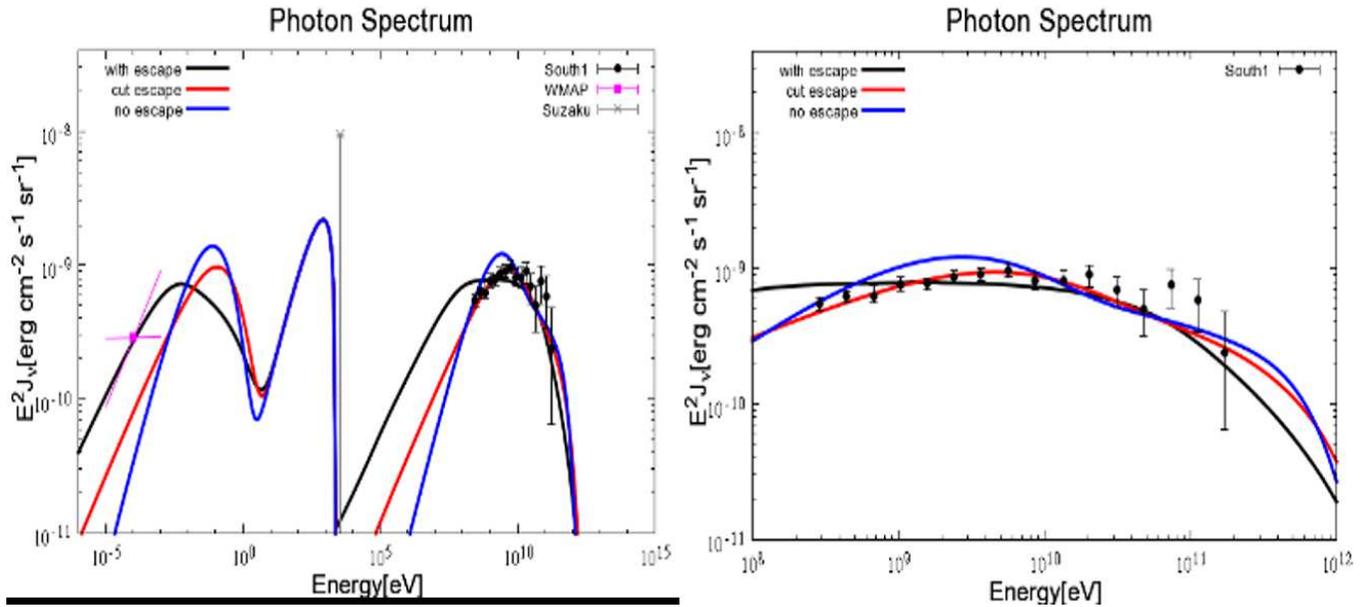}
\caption{{\textbf left:} Photon spectrum under three cases. {\textbf right:} Photon spectrum at gamma ray region. 'No escape model'(blue solid), 'Cut escape model'(red solid), 'With escape model'(black solid) are all shown.The data points are from \cite{yan14}.}
\label{fig:spectrum}
\end{figure*}

Our results are shown in the Figure~\ref{fig:spectrum}. We calculate equation (\ref{FPeq}) under the three cases at variable parameters $L$ and $Q_0$ for $t = 2 {\rm kpc}/v_{\rm pro} \simeq 8$ Myr. $v_{\rm pro}=250 {\rm km sec^{-1}}$ and $B=4 \mu$G are fixed.
The result of 'No escape model' is shown in blue solid line, 'Cut escape model' is shown in red solid line, and 'With escape model' is shown in black solid line. The data points in \cite{yan14} are adopted. 

Gamma ray data of the Fermi-LAT are nicely reproduced in all cases. But the microwave data of the WMAP are reproduced only by 'With escape model'(Figure~\ref{fig:spectrum}). This is because of increasing of low energy particles due to escape from the turbulence. In this case the turbulence size $L=10$ pc, smaller than that of in \cite{mer11} $L=2$ kpc.
 If we reproduce the WMAP data in other cases, stronger magnetic field is required at large scale.

\section{Summary}
\label{sec:sum}
We reproduce both the Fermi Bubble and the WMAP haze under typical galactic magnetic field $B=4 \mu$G by considering the effect of time dependence and electrons which escape from the turbulence. 'Cut escape model' which is similar to the model of previous research Mertsch \& Sarkar \cite{mer11} reproduce gamma ray data well, but predicted microwave emission is not enough at $B=4 \mu$G. Our results suggest that smaller turbulence size is better for reproducing microwave observation, because of more easily escaping from the turbulence. 

\acknowledgements

We appreciate JaCOW for helpful discussion.

\end{document}